\documentclass[a4paper,12pt,twocolumn]{article}
\usepackage{cite}
\usepackage[dvips,hiresbb]{graphicx}

\title{\textbf{Special polarization characteristic features of a three dimensional terahertz photonic crystal not apparently apply to physical and optical basic rules}}
\author{\Large{Chikara Sakurai} \thanks{Email: c-sakurai@river-ele.co.jp (primary), sakuraikazan@gmail.com (second)} \\ \textit{River Electric Corporation, 2-1-11 Fujimigaoka, Nirasaki, Yamanashi, Japan}} 
\date{\today}
\begin{document}
\onecolumn{
\maketitle
\begin{abstract}
\normalsize{A silicon inverse diamond structure whose lattice point shape was vacant regular octahedrons had a complete photonic band gap at around 0.4 THz and X point's photonic band gap (0.36 THz to 0.44 THz) by plane wave expansion method. 
It is said that three-dimensional photonic crystals have no polarization anisotropy in photonic band gap (stop gap, stop band) of high symmetry points in normal incidence.  Experimental results, however, confirmed that the polarization orientation (electric-field direction) of a reflected wave was different from that of an incident wave whose direction was [001].
The polarization orientation of the incident wave was parallel to the surface (001) of the photonic crystal, and it was set in the orientation defined as $\theta$ $^\circ$ (degree).  The angle, $\theta$ was 0$^\circ$ to 90$^\circ$ per 15$^\circ$. A sample was rotated in plane (001) instead of the incident wave, relatively.
The polarization orientation of the reflected wave was parallel to that of the incident wave for $\theta$ = 0$^\circ$ and 90$^\circ$, in contrast, the former was perpendicular to the latter for $\theta$ = 45$^\circ$ in the vicinity of 0.42 THz. For an intermediate $\theta$, the former was an intermediate orientation. As far as the photonic crystal in this work is concerned, these phenomena do not apply to physical and optical basic rules in appearance. } 
\end{abstract}

\twocolumn{
\section*{Introduction}
\hspace*{5mm} Terahertz (THz) waves are located halfway between microwaves and infrared lights. There are many materials that are opaque in the visible region but transparent in the THz region. Many crystals have specific absorption spectra in the THz region. Recently,  frequency applications from millimeter waves to THz waves have been steadily proceeding for a next-generation fast wireless communication.
In this way, the THz techniques allow non-destructive and non-contact observations, and  applications as imaging, sensors, identification of materials and the high-capacity wireless communication~\cite{kaw, ari, oda, hos, sal1, adr, jan, kou}.\\
\hspace*{5mm}The photonic crystals (PC) that have the regular periodicity of dielectric materials~\cite{yab, ozb, tak, nod, kwa, Lin, chi, Not}, are able to control waves whose sizes are comparable to their lattice constant by forming point defects and line defects (waveguide) on their surface and within them.  
The PC without forming above structures also indicate characteristic properties such as complete photonic band gap (CPB), negative refractive index, slow velocity and so on under certain conditions.\\
\hspace*{5mm}This work with the polarization anisot- ropy confirmed that new reflection phenomena of the three-dimensional (3D)-PC do not apparently apply to general physical and optic basic rules.\\
\hspace*{5mm}These basic rules are as follows. \\
\hspace*{5mm}(A)  Electric-field is vector, and various optical phenomena are analyzed by using methods of resolution and synthesis of one.\\
\hspace*{5mm}(B)  According to Maxwell's equations that are electromagnetic basic rules, 
the electric-field direction of the reflected wave rotates by 180 degrees from that of the incident wave in the case of transmittance zero and normal incidence.\\
\hspace*{5mm}(C)  According to crystallography, diamond structure is in cubic system and optically isotropic.
\\
\hspace*{5mm}In this work, more detail process and fabricating methods of sample, experimental data and analyses than those of ref.~\cite{sakurai} are explained. \\
\hspace*{5mm}The band structure of the 3D-PC in a terahertz (THz) region with the Si inverse diamond structure whose lattice point shape was vacant regular octahedrons, was calculated by using plane wave expansion method~\cite{sakurai}. These theoretical results confirmed that CPB existed at around 0.4 THz.\\
\hspace*{5mm}The polarization anisotropy, which was the polarization orientation (electric-field direction) difference between the reflected wave and the incident one, was studied on the surface (001) at around BGX that is X point's photonic band gap  (0.36 THz to 0.44 THz). The 3D-PC sample was rotated in plane (001) instead of the incident wave, relatively. The rotation angle $\theta$ was 0$^\circ$ to 90$^\circ$ per 15$^\circ$ (degrees).}\\
\hspace*{5mm}The polarization orientations of  TE and TM waves are orthogonal each other. The 2D-PC~\cite{JDJ} has different band structures for two waves, and it generally indicates different polarization features. In the 3D-PC, the polarization anisotropy of TE and TM waves exists with Brewster's angle on the surface,  internally located waveguide and so on in a frequency region~\cite{pri, lid}.  The 3D-PC has eigen modes in frequencies except the photonic band gap. Each eigen mode has an intrinsic symmetry, and some polarization anisotropy also exists~\cite{min}.\\
\hspace*{5mm}However, the polarization anisotropy in this work is entirely different from above characteristics. For example, no polarization anisotropy of two reflected waves whose polarization orientation is orthogonal each other in appearance
\footnote{It was found that a phase-difference exists between these two waves by the later FEM (finite element method) analyses, arXiv:1811.02990.}, exists in normal incidence and within the photonic band gap in which no eigen modes exist.  However, under above two conditions, the 3D-PC indicates a reflectance property such as the reflective 1/2 wave plate almost without energy loss in spite of the optically isotropic PC. The special polarization anisotropy exists as discussed below. 
\section*{Photonic Band Structure}
\hspace*{5mm}The lattice of the diamond structure is shown as fig.~\ref{fig:latticeA}(a). The sphere is the lattice point and its shape is the regular octahedrons in fig.~\ref{fig:latticeA}(b). 
It is vacant (atmosphere) and the dielectric constant, $\varepsilon_1 = 1.00$ was set. 
The surrounding material was pure Si (resistivity, $\rho >10^4\ \Omega$ cm) and the dielectric constant, $\varepsilon_2 = 11.9$ was set.
The lattice constant, $a = 300\ \mu$m and the length of the regular octahedrons side,
$L = 150\ \mu$m were set in this theoretical and experimental work.  In the Si cube lattice, many sections of the air octahedrons on the lattice points are arranged as shown in 
fig.~\ref{fig:latticeA}(c). \\
\begin{figure}[h]
\centering
\includegraphics[width=7cm,]{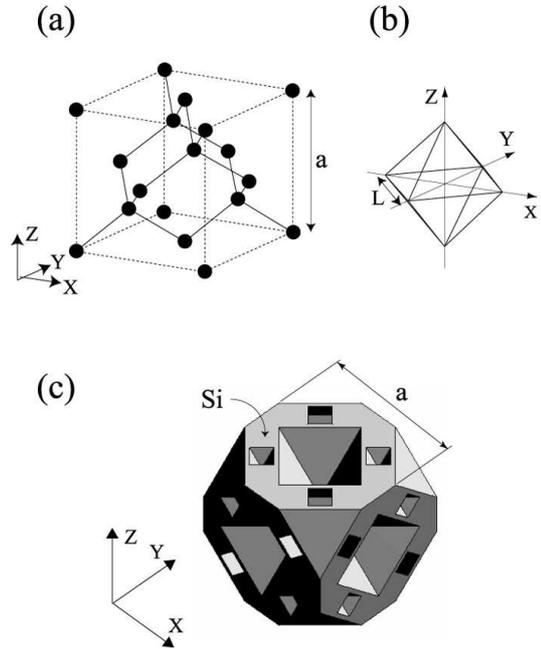}
\caption{\footnotesize (a) Lattice of the diamond structure.  (b) Shape of the lattice point is regular octahedrons. It is vacant ($\varepsilon_1$ = 1.00). The surrounding material is Si  ($\varepsilon_2$ = 11.9). (c) Si cube lattice and many sections of the air octahedrons on the lattice points.}
\label{fig:latticeA}
\end{figure}
\hspace*{5mm}Fig.~\ref{fig:bandstructure}(a) shows the calculated photonic band structure by using plane wave expansion method and CPB exists at around 0.4 THz.
The first Brillouin zone is shown as fig.~\ref{fig:bandstructure}(b).
In this work, the direction of the incident wave was [001]
\footnote{[001], (001), and \{100\} etc. are defined on the X-Y-Z coordinate system in 
fig.~\ref{fig:latticeA}(a).}
in the real space and it corresponds to $\Gamma$-X direction in the wave number space (K-space). BGX exists between 0.36 THz and 0.44 THz.
\begin{figure}[h]
\centering
\includegraphics[width=7.8cm,]{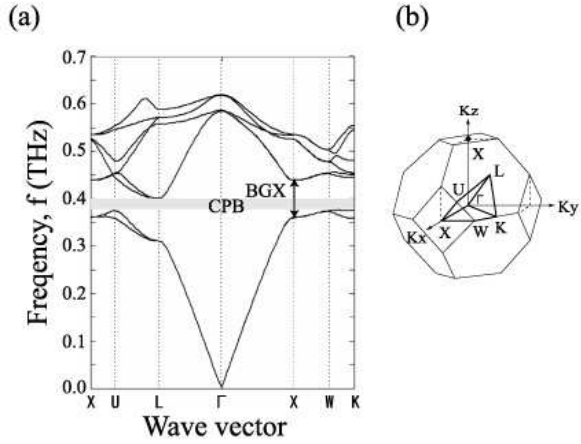}
\caption{\footnotesize (a) Calculated photonic band structure has CPB (gray zone) at around 0.4 THz. BGX exists between 0.36 THz and 0.44 THz. (b) First Brillouin zone and the reduced zone (heavy line) with high symmetry points.}
\label{fig:bandstructure}
\end{figure}
%
\section*{Process and Fabrication}
\hspace*{5mm}Fig.~\ref{fig:picture} shows the 3D-PC sample layered and fabricated. X-Y-Z coordinate system corresponds to that in fig.~\ref{fig:latticeA}. The total number of layered chips was 28 (height: 7a).\\
\hspace*{5mm}The main periodic square patterns (side $L = 150\ \mu$m) on both surfaces (001) of Si chip, whose area was $10\times 10$ mm$^2$ and whose height was $75\ \mu \ 
\mathrm{m} (= a/4)$, were etched by a normal method, and they were periodically arranged along the direction [001]. \\
\hspace*{5mm}Strong alkaline solution, KOH was used for etching. The etching time depends on the Si wafer's thickness, a temperature and a concentration of the liquid solution and so on. In this work, KOH concentration was 40 wt\%, the temperature was about 80$^\circ$C, and the etching time was about 5 hours.\\
\hspace*{5mm}A frame was set around each Si chip. V channels on the front and rear surface of the frame were etched and formed for fixing adjacent chips. Adhesive agent and Si balls were put in the space of two V channels whose cross-section was rhombic. Clearance grooves on the frame for impounding extra adhesive agent were formed on the inner side of the V channels.\\
\begin{figure}[t]
\centering
\includegraphics[width=7cm,]{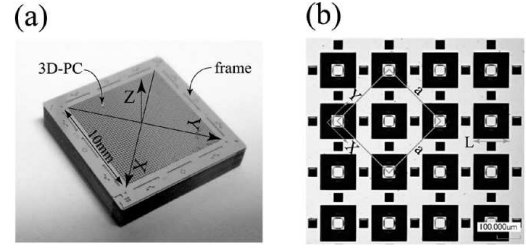}
\caption{\footnotesize (a) Stereograph of the 3D-PC sample with the Si inverse diamond structure.   (b) Surface picture viewed from directly above.}
\label{fig:picture}
\end{figure}
%
\begin{figure}[t]
\centering
\includegraphics[width=8.2cm,]{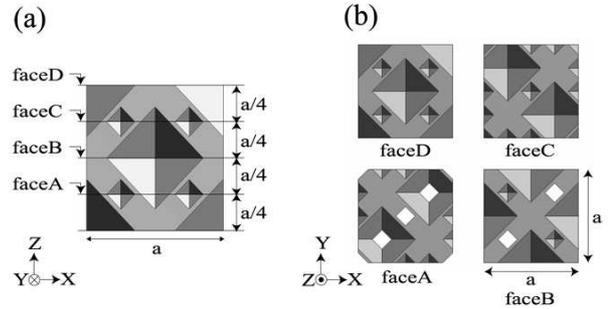}
\caption{\footnotesize (a) Lattice viewed from Y-direction with each layer surface name, face A to D. (b) Face A to D viewed from Z-direction.  Face D corresponds to a $\times$ a area in fig.~\ref{fig:picture}(b).}
\label{fig:latticeB}
\end{figure}
\hspace*{3mm}The etching angle of Si \{100\} surface is $54.7^{\ \circ}$ and the etched vacant shape forms the regular octahedrons between four layers.  The lattice of the Si inverse diamond structure in fig.~\ref{fig:latticeA}(c) viewed from Y-direction is shown in 
fig.~\ref{fig:latticeB}(a).  The vacant regular octahedrons located in the center, one-half of which appears in the figure, are divided into four parts (two small square pyramids and two large truncated square pyramids~) in the Z-direction. The patterns of these small square pyramids appear as small square ones in fig.~\ref{fig:picture}(b).
    Face A to D is the topside surface of each layer. Fig.~\ref{fig:latticeB}(b) shows face A to D viewed from Z-direction.
%
\section*{Experimental System}
\hspace*{5mm}Fig.~\ref{fig:system} shows the schematic diagram of the measurement system with the THz-TDS (time-domain spectroscopy) equipment owned by Nippo Precision co., ltd.. The coordinate system, x-y-z is used as that of this measurement system.\\
\begin{figure}
\centering
\includegraphics[width=7.5cm,]{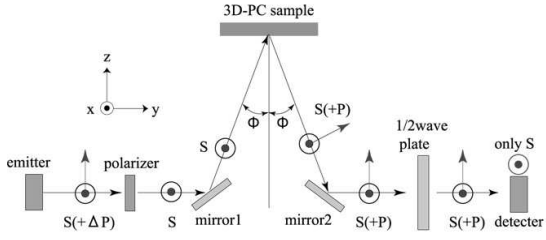}
\caption{\footnotesize Schematic diagram of the measurement system with the THz-TDS. The 3D-PC sample is set as the layered direction $\|$ z-axis. The polarization orientation of a THz wave from the emitter is S-p $\|$ x-axis and the detector detects only S-p. The designed 1/2 wave plate converts S-p into P-p and P-p into S-p at around 0.42 THz. S(+P) means S-p, P-p or the mixing of S-p and P-p.}
\label{fig:system}
\end{figure}
\hspace*{5mm}The polarization orientation of a THz wave from the emitter is parallel to x-axis and it is called S-polarization (S-p).
The mirror 1 and mirror 2 are coated with Au.
S-p launched by the emitter is reflected by the mirror 1 and the orientation of the polarization is also S-p 
\footnote{The expression of the inversion, phase shifting by 180$^\circ$, of S-p is abbreviated in fig.~\ref{fig:system}.}.
The polarizer is used for obtaining more precise S-p. 
The 3D-PC sample is so horizontally set that the layered direction is parallel to z-axis; it is
parallel to Z-axis in fig.~\ref{fig:latticeA}(a). The incident angle, $\phi$ is $7^{\ \circ}$, which is the normal incidence approximately.\\
\hspace*{5mm}Another polarization perpendicular to S-p is P-polarization (P-p); it is included in the incidence plane $\|$ z-axis.
According to Fermat's principle, the reflection angle is equal to the incident one. When not only S-p but P-p is included in the reflected wave of the 3D-PC sample, S-p and P-p are naturally included in the reflected wave by mirror 2. Meanwhile, the detector detects only S-p.
A 1/2 wave plate is used for the measurement of a P-p component included in the reflected wave of the 3D-PC sample.\\
\begin{figure}
\centering
\includegraphics[width=4cm,]{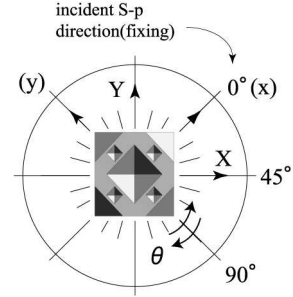}
\caption{\footnotesize Explanatory diagram of the 3D-PC sample's rotation angle, $\theta$ (degree). X- and Y-axis are fixed on the 3D-PC sample. They rotate with the sample around z-axis. The incident S-p direction is fixed at 0$^\circ$. 0$^\circ$ is fixed and it corresponds to x-axis in fig.~\ref{fig:system}. $\theta$ is defined as 0$^\circ$ when X-axis is parallel to 45$^\circ$. For example, $\theta$ is 45$^\circ$ when X-axis is parallel to 0$^\circ$.}
\label{fig:rotation}
\end{figure}
\hspace*{5mm}In this work, the 3D-PC sample was rotated in plane (001) instead of the S-p incident wave, relatively.  Fig.~\ref{fig:rotation} shows the explanatory diagram of the 3D-PC sample's rotation angle, $\theta$ (degree). X- and Y-axis fixed at the 3D-PC sample correspond to the coordinate system in fig.~\ref{fig:latticeA}; two axes are also rotated with the sample.  x- and y-axis are not rotated and it is fixed at the measurement system in fig.~\ref{fig:system}. The incident S-p direction is fixed at 0$^\circ$ $\|$ x-axis.  Scale marks, 0$^\circ$, 45$^\circ$
and 90$^\circ$ are also fixed. $\theta$ is defined as 0$^\circ$ when the 3D-PC sample is located as shown in fig.~\ref{fig:rotation}; X-axis is parallel to 45$^\circ$. For example, $\theta$ is 45$^\circ$ when X-axis is parallel to 0$^\circ$. \\
\hspace*{5mm}The material of the 1/2 wave plate is quartz SiO$_2$. Xc and Zc (crystal axes) are included in the plane of the plate. Yc is parallel to the direction of the thickness; it is parallel to y-axis in fig.~\ref{fig:system}.  The ordinary and extraordinary refractive indices are $n_\mathrm{o} = n_\mathrm{Xc} = n_\mathrm{Yc} =  2.108$
and $n_\mathrm{e} = n_\mathrm{Zc} = 2.156$ at 1 THz~\cite{JBG}, respectively.\\
\begin{figure}[t]
\centering
\includegraphics[width=4cm,]{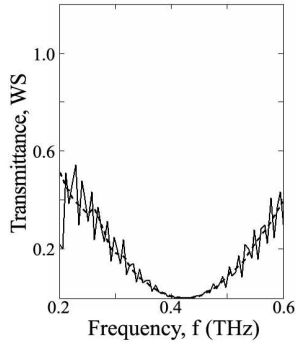}
\caption{\footnotesize Transmittance, $WS(f)$ of the designed 1/2 wave plate. For example, S-p incident wave is nearly converted into P-p at 0.42 THz. The dash line includes the signal disposal from the back surface of the 1/2 wave plate. The solid line includes no disposal. }
\label{fig:WS}
\end{figure}
\hspace*{5mm}The designed 1/2 wave plate whose thickness is 7.50 mm converts S-p into P-p and P-p into S-p at around 0.42 THz when one of the two bisector of Xc and Zc is parallel to x-axis.
Meanwhile, when either Xc or Zc is parallel to x-axis, the orientations of S-p and P-p remain unchanged after passing through  the 1/2 wave plate. \\
\hspace*{5mm}The average of the transmitted spectra of S-p $\|$ Xc and S-p $\|$ Zc, was  defined as $avXZ(f)$; it was used for the normalization of the characteristic features of the 1/2 wave plate. The variable, $f$ is a THz frequency.\\
\hspace*{5mm}Fig.~\ref{fig:WS} shows transparent characteristics, $WS(f)$
of the 1/2 wave plate, which is the spectrum normalized by $avXZ(f)$ from 0.2 to 0.6 THz . $WS(f)$ is at a minimum at 0.42 THz within BGX  (0.36 THz to 0.44 THz).
$1-WS(f)$ is the ratio of P-p conversion into S-p, inversely.\\
\hspace*{5mm}The solid line includes the concavity and convexity: the reflection of the back surface of the 1/2 wave plate. In electric-field function, $E(t)$~(t: time) before Fourier transform, a second peak appeared after a main peak group that was signal from the 3D-PC sample. The delay time was nearly equal to round-trip time of the  1/2 wave plate. The dash line is the spectrum which deletes this second peak.\\
\hspace*{5mm}In this work, the solid line with no disposal was used. 
In measurement of $WS(f)$ and $avXZ(f)$, the incidence wave was S-p and an Au-plate was set instead of the 3D-PC sample.  
\section*{Experimental Results}
\begin{figure}[t]
\centering
\includegraphics[width=6cm,]{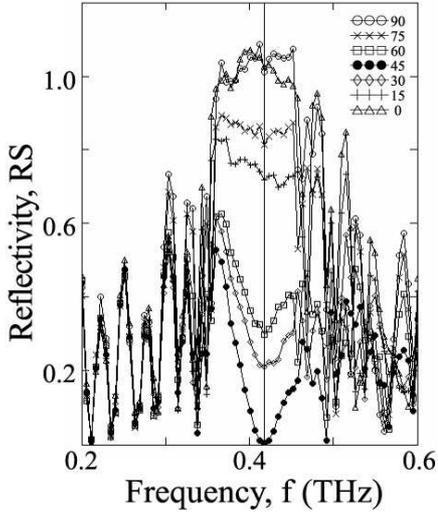}
\caption{\footnotesize Reflection spectra, $RS(f)$'s measured with no 1/2 wave plate from 0.2 to 0.6 THz for $\theta$ = 0$^\circ$ to 90$^\circ$ per 15$^\circ$.  The vertical solid line is 0.42 THz. Seven kinds of $RS(f)$ consist of only S-p. For $\theta$ = 0$^\circ$ and 90$^\circ$, $RS(f)$ is almost 1.0 at around BGX. Meanwhile,
For $\theta$ = 45$^\circ$, $RS(f)$ is nearly equal to zero especially at 0.42 THz.}
\label{fig:RS}
\end{figure}
\begin{figure}[h]
\centering
\includegraphics[width=7.9cm,]{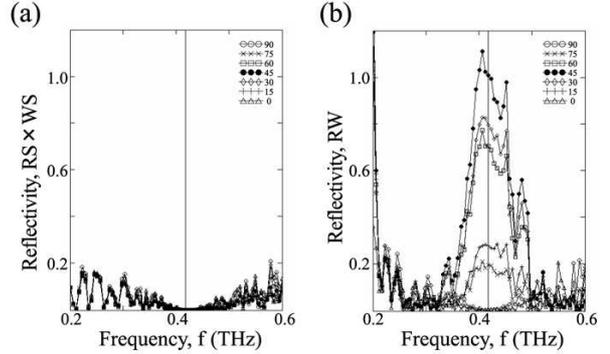}
\caption{\footnotesize (a) $RS(f)\times WS(f)$ is the reflected spectrum for each $\theta$ when only S-p passes through the 1/2 wave plate. (b) $RW(f)$ is the reflected spectrum for each $\theta$, which passes through the 1/2 wave plate and it is normalized by $avXZ(f)$.} 
\label{fig:RWRSWS}
\end{figure}
%
\begin{figure}[t]
\centering
\includegraphics[width=6cm,]{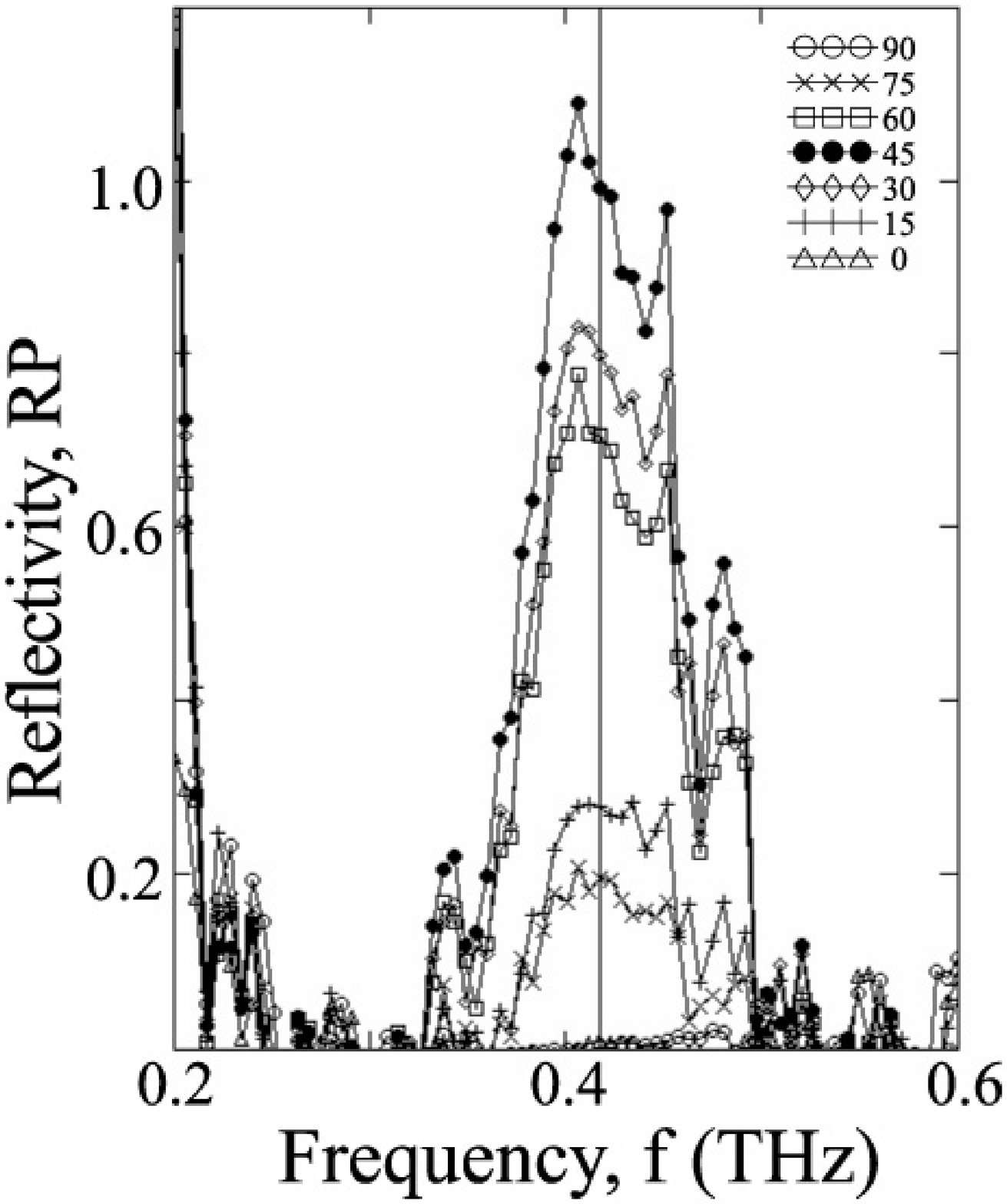}
\caption{\footnotesize $RP(f)$'s are the normalized reflected spectra including only P-p component.
 $RP(f)$ is $[RW(f)-RS(f)\times WS(f)] /[1-WS(f)]$.}
\label{fig:RP}
\end{figure}
\hspace*{5mm}Fig.~\ref{fig:RS} shows the S-p reflected spectra,~seven kinds of $RS(f)$ measured at around BGX. $RS(f)$ is the reflected spectrum with no 1/2 wave plate and it is normalized by the Au reflected one. The rotation angle, $\theta$ is 0$^\circ$ to 90$^\circ$ per 15$^\circ$. Seven kinds of $RS(f)$ consist of only S-p. For $\theta$ = 0$^\circ$ and 90$^\circ$, $RS(f)$ is almost 1.0 at around BGX. Meanwhile,
For $\theta$ = 45$^\circ$, $RS(f)$ is nearly equal to zero especially at 0.42 THz. For other $\theta$'s (15$^\circ$, 30$^\circ$, 60$^\circ$, and 75$^\circ$), A series of $RS(f)$ indicate intermediate values. \\
\hspace*{5mm}Fig.~\ref{fig:RWRSWS} shows $RS(f)\times WS(f)$ and $RW(f)$ for each $\theta$.
$[RS(f)\times WS(f)]$'s are the reflected spectra when only S-p passes through the 1/2 wave plate.
$RW(f)$'s are the reflected ones passing through the 1/2 wave plate and it is normalized by $avXZ(f)$.\\
\hspace*{5mm}Seven kinds of $RS(f)\times WS(f)$ for all $\theta$ (0$^\circ$ to 90$^\circ$ per 15$^\circ$) indicate similar frequency dependence since $WS(f)$ is convex downward and small at around BGX.\\
\hspace*{5mm}$RS(f)\times WS(f)$ is nearly equal to $RW(f)$ for $\theta$ = 0$^\circ$ and 90$^\circ$. In other words, the reflected spectra for $\theta$ = 0$^\circ$ and 90$^\circ$
consist of only S-p .\\
\hspace*{5mm}Meanwhile, for $\theta$ = 45$^\circ$, $RS(f)\times WS(f)$ is entirely different from $RW(f)$ especially at 0.42 THz. $RS(f)\times WS(f)$ is nearly equal to zero. In contrast, $RW(f)$ is almost 1.0 at 0.42 THz. In other words, for $\theta$ = 45$^\circ$, the reflected spectrum consists of almost P-p especially at 0.42 THz
 \footnote{The reflected spectra for $\theta$ = 135$^\circ$ have the same properties as those for $\theta$ = 45$^\circ$ in fig.~\ref{fig:RS} to fig.~\ref{fig:RP}.}.\\
\hspace*{5mm}For other $\theta$ (15$^\circ$, 30$^\circ$, 60$^\circ$, and 75$^\circ$), the reflected spectra include both S-p and P-p  at around BGX.\\
\hspace*{3mm} $RW(f)$ is not exact P-p~spectrum;~it~includes~S-p~component. $[RW(f)-RS(f)\times WS(f)]$ is the spectrum with the exception of  S-p component. Moreover, 
except the influence of the 1/2 wave plate, the normalized reflected spectrum with only P~-~p ~component, $RP(f)$ is $[RW(f)-RS(f)\times WS(f)]/[1-WS(f)]$, which is shown in 
fig.~\ref{fig:RP}.\\
\hspace*{5mm}S-p incident wave for $\theta$ = 45$^\circ$ is almost entirely converted into P-p reflected wave especially at around 0.42 THz
\footnote{$R10(f)$ is minimum, and $\Delta R10(f)/[1-WS(f)]$ is maximum, at around 0.38~THz in fig.~7, ref.~\cite{sakurai}. In this work, the process and fabrication were improved, and the polarizer was used as shown in fig.~\ref{fig:system}. It is likely that the difference in two frequency values depends on them.}.
Meanwhile, for $\theta$ = 0$^\circ$ and 90$^\circ$, P-p component
is nearly equal to zero. For other $\theta$, A series of $RP(f)$ indicate intermediate values.
The magnitude relation between $RP(f)$ and $RS(f)$ for seven $\theta$ reverses at 0.42 THz.\\
\hspace*{5mm}The angle, $\alpha$$_{sp}$ ($\theta$) is the one between P-p and S-p at 0.42 THz, which is shown in fig.~\ref{fig:Asp}(b). \\
\hspace*{5mm}$tan$($\alpha$$_{sp}$) = $\sqrt{RP}$/$\sqrt{RS}$ (0$^\circ$ $\le$ $\alpha$$_{sp}$ $\le$ 90$^\circ$) \\
$W (\psi)$ is the theoretical angle between two polarization waves just before and after passing through the 1/2 wave plate.\\
\hspace*{5mm}In fig.~\ref{fig:Asp}(c), Xc and Zc is the crystal axes of quartz.
 The refractive index, $n_\mathrm{Zc}$ is larger than $n_\mathrm{Xc}$.
\textit{OA} is a polarization vector just before passing through the 1/2 wave plate, and \textit{OB} is a polarization vector after passing through the 1/2 wave plate.  
\textit{OC} is the inverse vector of \textit{OB}.
The variable, $\psi$ is the angle between Zc-axis and \textit{OA}.\\
\hspace*{5mm}For the only intensity measurement, \textit{OB} and \textit{OC} are indistinguishable. Therefore, $W$ is defined as min($|W_1|, |W_2|$) from 0$^\circ$ to 90$^\circ$.
$W_1$ and $W_2$ are shown in fig.~\ref{fig:Asp}(c).\\
\hspace*{5mm}Characteristic features of  $\alpha$$_{sp}$ ($\theta$) and $W (\psi)$ are shown in fig.~\ref{fig:Asp}(a). $W (\psi)$ is the theoretical curve.
$\alpha$$_{sp}$ ($\theta$) was nearly equal to $W (\psi)$ for the intensity.
These results confirmed that the 3D-PC in this work would be available as the reflective 1/2 wave plate almost without energy loss.
\begin{figure}[t]
\centering
\includegraphics[width=7.5cm,]{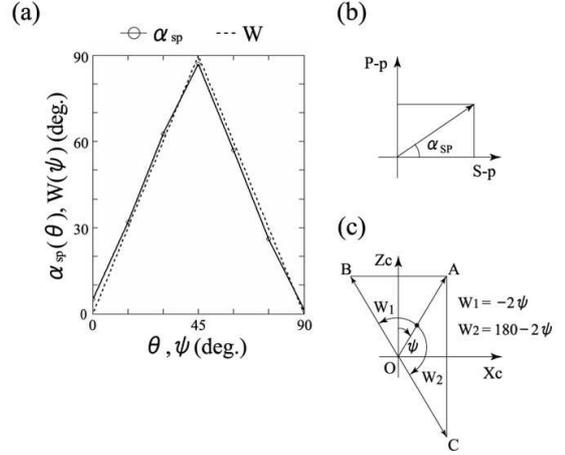}
\caption{\footnotesize (a) Solid line: $\alpha$$_{sp}$ ($\theta$), angle between P-p and S-p. dashed line: $W (\psi)$, theoretical angle between two polarization waves just before and after passing through the 1/2 wave plate. $\theta$ is the rotation angle in fig.~\ref{fig:rotation}. (b) Explanatory diagram of $\alpha$$_{sp}$. (c) Explanatory diagram of $\psi$, $W_1$ and $W_2$.}
\label{fig:Asp}
\end{figure}
\section*{Discussions}
\hspace*{5mm}The theoretical analyses confirmed that the Si inverse diamond structure (lattice constant, $a = 300$ $\mu$m) whose lattice point shape was vacant regular octahedrons (side, $L = 150$ $\mu$m), had CPB at around 0.4 THz within BGX that is X point's photonic band gap~\cite{sakurai}. The special polarization characteristic features were measured with the conditions as follows. Firstly, the incident wave was normal incidence; its direction was [001] ($\Gamma$-X direction) approximately. Secondly, the polarization orientation of the incident wave was parallel to the surface (001) of the 3D-PC sample. Thirdly, the sample was relatively rotated instead of the incident wave in the plane (001) from 0$^\circ$  to 90$^\circ$ per 15$^\circ$.\\
\hspace*{5mm}The experimental results confirmed that the polarization orientation of the reflected wave was different from that of the incident wave especially at around 0.42 THz within BGX in which no eigen modes exist.   \\
\hspace*{5mm}Typical four polarization characteristic features at 0.42 THz are shown in fig.~\ref{fig:conclusion}. Four polarization orientations of the incident wave are [X], [Y], [~Y~ =~X~]
and [Y~=~$-$~X], which correspond to 45$^\circ$, 135$^\circ$, 90$^\circ$ and 0$^\circ$
for $\theta$, respectively.\\
\hspace*{5mm}In the case of the polarization orientation of the incident wave $\|$ [~Y~ =~X~] \{ or [Y = $-$X]\} direction, that of the reflected wave was parallel to [~Y~ =~X~] \{ or [Y = $-$X]\} direction,  respectively.
However, in the case of that of the incident wave $\|$ [X] \{or [Y]\} direction, that of the reflected wave was parallel to [Y] \{or [X]\} direction,  respectively.\\
\hspace*{5mm}In addition, in the case of other incident wave's polarization, each angle of two polarization orientations between the incident wave and the reflected one was an intermediate value above 0$^\circ$ below 90$^\circ$, respectively.\\
\hspace*{5mm}The almost complete S-P transformation appeared at around 0.42 THz within BGX almost without energy loss.
These characteristic features indicate that the 3D-PC in this work will be available as the reflective 1/2 wave plate.\\
\begin{figure}[t]
\centering
\includegraphics[width=6cm,]{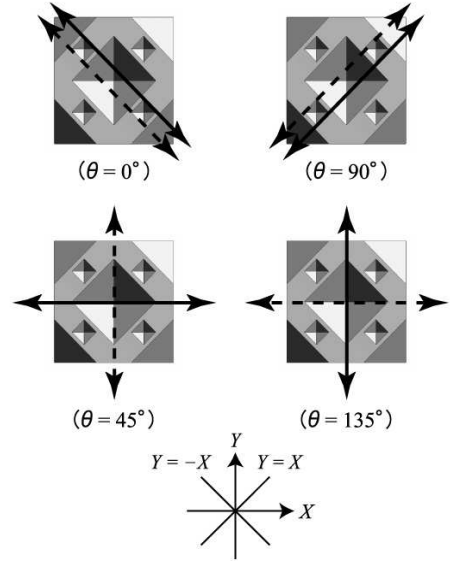}
\caption{\footnotesize Typical four polarization characteristic features at 0.42 THz.
In the case of the polarization orientation of the incident wave (solid arrow) $\|$ [Y~ =~X] \{or [Y~=~$-$~X]\} direction, that of the reflected wave (dash arrow) is parallel to [Y~ =~X] \{or [~Y~=~$-$~X~]\} direction.
However, in the case of that of the incident wave $\|$ [X] \{or [Y]\}, that of the reflected wave is parallel to [Y] \{or [X]\}.}
\label{fig:conclusion}
\end{figure}
\hspace*{5mm}These results above do not apparently apply to basic rules (B) and (C).\\
\hspace*{5mm}The two directions of [Y~ =~X] and [Y~=~$-$~X] are constitutionally and optically comparable. According to rule (A), in the case of the polarization orientation of the incident wave $\|$ [X] \{or [Y]\}, that of the reflected wave is expected to be parallel to [X]  \{or [Y]\}, respectively by the synthesis of two incident polarization vectors, {[~Y~ =~ X~] and [Y~=~$-$~X]}.
However, experimental results indicated that the polarization orientation of the incident wave and that of the reflected wave bisected at right angles each other at around 0.42 THz. \\
\hspace*{5mm}These results above do not apparently apply to basic rule (A)
\footnote{In footnote 1, the existence of the phase difference was confirmed, however, it still remains unsolved that the reason why the polarization orientation of the reflected wave is rotated by 90$^\circ$ for that of the incident wave.}.\\
\section*{Conclusion}
\hspace*{5mm}As far as the polarization characteristic features of the 3D-PC in this work is concerned, physical and optical basic rules do not apply to these experimental results in appearance.
\section*{Acknowledgment}
\hspace*{5mm}The author would like to thank Hidekazu
Onishi, who fabricated the designed Si inverse
diamond structure and Ph.D. Takeshi Sawada,
who is a research scientist, Terahertz Project,
Second Design Department, Nippo Precision
co., ltd..
\\

}

\begin{thebibliography}{99}
 \bibitem{kaw}
 K.~Kawase, Optics and Photonics News 15, 34 (2004).
 
 \bibitem{ari}
 S.~Ariyoshi, C.~Otani, A.~Dobroiu, H.~Sato, K.~Kawase, H.~M.~Shimizu, T.~Taino and H.~Matsuo, Appl. Phys. Lett. 88 203503 (2006).
 
 \bibitem{oda}
 N.~Oda, A.~W.~M.~Lee, T.~Ishi, I.~Hosako and Q.~Hu, Proceedings of the SPIE, 8363 (2012).
 
 \bibitem{hos}
 H.~Hoshina, Y.~Sasaki, A.~Hayashi, C.~Otani and K.~Kawase. Appl. Spectrosc., 63, 81, (2009).
 
 \bibitem{sal1}
 K.~A.~Salek, H.~Nakanishi, A.~Ito, I.~Kawayama, H.~Murakami and M.~Tonouchi, Opt. Eng. 53 (3) 031204 (2013).

 \bibitem{adr}
 A.~Dobroiu,~M.~Yamashita, Y.~N. Ohshima, Y.~Morita, C.~Otani, and K.~Kawase, Applied Optics, 43, 30, (2004) 5637.

 \bibitem{jan}
 J.~Hebling, G.~Almasi, I.~Z.~Kozma, and J.~Kuhi, 
Optics Express, 10, 21 (2002) 1161.

 \bibitem{kou}
 K.~Nawata, T.~Taira, J.~Shikata, K.~Kawase and H.~Minamide,
Scientific Reports 4, 5045 (2014).

 \bibitem{yab}
 E.~Yablonovitch, T.~J.~Gmitter and K.~M.~Leung, Phys.~Rev.~Lett.~67 (1991)~2295.
 
 \bibitem{ozb}
 E.~Ozbay, E.~Michel, G.~Tuttle, R.~Biswas, M.~Sigalas and K.~M.~Ho, Appl. Phys. Lett. 64 (1994) 2059.
 
 \bibitem{tak}
 K.~Takagi and A.~Kawasaki, Appl. Phys. Lett. 94 (2009) 021110.
 
 \bibitem{nod}
 S.~Noda, K.~Tomoda, N.~Yamamoto and A.~Chutinan, Science 289 (2000) 604.
 
 \bibitem{kwa}
 S.~Kawakami, T.~Kawashima and T.~Sato, Appl. Phys. Lett. 74 (1999) 463.
  
 \bibitem{Lin}
 Shawn-Yu~Lin, V.~M.~Hietala,\\
 Li~Wang, and E.~D.~Jones,
 Opt. Lett. 21, (1996) 1771. 
 
 \bibitem{chi}
 Chiyan~Luo, S.~G.~Johnson, and\\ 
 J.~D.~Joannopoulos, Appl. Phys. Lett. 81 (2002).
 2352

 \bibitem{Not} 
 Notomi~M., Yamada~K., Shinya~A., Takahashi~J., Takahashi~C. and Yokohama~I., Phys. Rev. Lett. 87, 253902 (2001). 
 
 \bibitem{sakurai}
 C.~Sakurai, arXiv:1703.08297
 [physics. optics] (2017).

 \bibitem{JDJ}
 J.~D.~Joannopoulos, S.~G.~Jhonson,\\
 R.~D.~Meade and
 J.~N.~Winn, Photonic Crystals, Princeton University Press
 (Printceton) (1995) 66-93.
 
 \bibitem{pri}
 Priya and R.~V.~Nair, Department of Physics, Indian Institute of Technology (IIT),
 Ropar Rupnagar, Punjab 140 001 INDIA.
 
 \bibitem{lid}
 E.~Lidorikis, M.~L.~Povinelli,
S.~G.\\Johnson, and J.~D.~Joannopoulos,  Physical Review Letters, 91, 2, 023902 (2003).
 
 \bibitem{min}
 Ming~Che, Zhi-Yuan~Li, and Rong-Juan~Liu, Phys.~Rev.~A~76,~023809 (2007).
 
 \bibitem{JBG}
 J-B.~Masson and G.~Gallot, Opt. Lett 265, 31,2 (2006).
 
\end{thebibliography}
\end{document}